\documentclass[prd,twocolumn,showpacs,amsmath,amssymb,footinbib]{revtex4}
\everymath{\displaystyle}
\usepackage[dvips]{graphicx}
\usepackage{longtable}
\newcommand \beq{\begin{eqnarray}}
\newcommand \eeq{\end{eqnarray}}

\def\qs{q\!\!\!/}

\def\simge{\mathrel{%
       \rlap{\raise 0.511ex \hbox{$>$}}{\lower 0.511ex \hbox{$\sim$}}}}
\def\simle{\mathrel{
       \rlap{\raise 0.511ex \hbox{$<$}}{\lower 0.511ex \hbox{$\sim$}}}}
\begin{document}

\title{Thermal Phase Transitions and Gapless Quark Spectra\\
in Quark Matter at High Density}
\author{K. Iida,$^{1,2}$  T. Matsuura,$^3$  
 M. Tachibana,$^2$ and T. Hatsuda$^3$}
\affiliation{$^{1}$
RIKEN BNL Research Center, Brookhaven National 
Laboratory, Upton, NY 11973\\
$^{2}$
The Institute of Physical and Chemical Research (RIKEN),  
 Wako, Saitama 351-0198, Japan\\
$^{3}$Department of Physics, University of Tokyo,
  Tokyo 113-0033, Japan}

\begin{abstract}
    Thermal color superconducting phase transitions in three-flavor quark
matter at high baryon density are investigated in the Ginzburg-Landau (GL) 
approach.  We constructed the GL potential near the boundary with a normal
phase by taking into account nonzero quark masses, electric charge 
neutrality, and color charge neutrality.  We found that the density of states 
averaged over paired quarks plays a crucial role in determining the phases 
near the boundary.  By performing a weak coupling calculation of the 
parameters characterizing the GL potential terms of second order in the 
pairing gap, we show that three successive second-order phase transitions 
take place as the temperature increases: a modified color-flavor locked 
phase ($ud$, $ds$, and $us$ pairings) $\to$ a ``dSC'' phase ($ud$ and 
$ds$ pairings) $\to$ an isoscalar pairing phase ($ud$ pairing) $\to$ a normal 
phase (no pairing).  The Meissner masses of the gluons and the number of 
gapless quark modes are also studied analytically in each of these phases. 
\end{abstract}
\maketitle

\section{Introduction}

     Over the past few years, the properties of color superconducting 
quark matter and the phase structure at high baryon density, which were 
originally studied in earlier publications \cite{BL}, have been 
examined intensively (see, e.g., Ref.\ \cite{RWA} for reviews).
A color superconductor predicted to occur in weak coupling is a relativistic 
system in which the long-range color magnetic interaction, which is 
screened only dynamically by Landau-damping of exchanged gluons, is 
responsible for the formation of the superconducting gap.  The gap in weak
coupling has a non-BCS form, 
 $\Delta \propto \mu \exp( -3 \pi^2/\sqrt{2} g )$
with $g$ the strong coupling constant and $\mu$ the quark chemical potential
\cite{SON99}.  Furthermore, the gap has a matrix structure with 
$(N_c \times N_f)^2$ components due to various combinations of $N_c$ color 
and $N_f$ flavor degrees of freedom.

     There is a good deal of evidence, obtained from various weak coupling  
analyses \cite{Sch-NPB-575,Evans-NPB-581,IB,HH-PRD-68}, that  
if all the quark masses are zero ($m_u=m_d=m_s=0$) and hence the quark 
chemical potentials are equal ($\mu_u=\mu_d=\mu_s$), the quark matter is 
in a color-flavor locked (CFL) phase \cite{CFL} at low temperature. 
A transition from the CFL phase to the normal phase in mean-field theory is 
of second order.  In weak coupling, the transition temperature $T_c$ is
related to the zero temperature gap $\Delta_{T=0}$ as 
$T_c= 2^{1/3} \times 0.57 \Delta_{T=0}$
\cite{SWR}, which is a BCS relation 
except for a factor $2^{1/3}$.

     The massless situation thus described may be approximately realized at 
asymptotically high densities where the quark masses are negligible compared 
to the quark chemical potentials.  However, the effect of nonzero quark 
masses becomes important when the chemical potentials decrease.  In the 
presence of a quark mass difference characterized by $2m_s/(m_u+m_d) \sim 25$,
$\beta$ equilibrium, electric neutrality, and color neutrality combine to 
produce non-trivial chemical potential differences between flavors and 
and between colors, leading to new phases such as the gapless CFL (gCFL) 
phase \cite{gCFL} in which two out of nine quark quasiparticles are gapless.

     In our recent Letter \cite{IMTH04}, we have studied what kind of phase 
structure appears in $\beta$ equilibrated neutral quark matter near the 
super-normal phase boundary when there are quark mass differences.
An advantage of studying the region near the phase boundary is that 
in classifying the possible phase structures we can make use of the 
model-independent Ginzburg-Landau (GL) analysis \cite{IB,IB3} in which
the thermodynamic potential difference between the superfluid and normal 
phases is expanded in terms of the order parameter (the pairing gap). 
Furthermore, by calculating the parameters controlling the thermodynamic 
potential terms in weak coupling, one can {\rm prove} which phase is 
realized below the super-normal boundary in the high density regime 
where the weak coupling analysis is valid.
      
      A crucial observation found in Ref.\ \cite{IMTH04} is that the phase 
structure near the critical temperature is essentially dictated by the 
average density of states of different quarks.  We also found that
the quark mass difference and the electric charge neutrality
(but not the color neutrality) play key roles in determining the phase 
structure near the critical temperature.  In weak coupling, it was shown that
the following successive phase transitions occur near the super-normal
phase boundary: a modified color-flavor locked (mCFL) phase ($ud$, $ds$, 
and $us$ pairings) $\to$ a ``dSC'' phase ($ud$ and $ds$ pairings) $\to$ an 
isoscalar two-flavor superconducting (2SC) phase ($ud$ pairing) $\to$ a 
normal phase (no pairing). 
 
      The purposes of this paper are (i) to give a detailed account of the 
results given in Ref.\ \cite{IMTH04} and (ii) to investigate the elementary 
excitation modes (gluons with the Meissner mass, massless gluons, gapped 
quarks, and gapless quarks).
  
      The content of this paper is as follows.  In Sec. II, we start with  
a toy model that captures the essential features of the hierarchical 
structure of the phase transitions.  Then we construct a general form of the 
GL potential with the quark masses in such a way that it fulfills symmetry 
constraints.  In Sec.\ III, we evaluate the parameters characterizing the GL 
potential terms in the weak coupling region with unequal quark masses 
($m_{i}$, $i$=$u, d, s$) and unequal quark chemical potentials ($\mu_{i}$, 
$i$=$u, d, s$) by utilizing the Cornwall-Jackiw-Tomboulis (CJT) effective 
action \cite{CJT}.  In Sec.\ IV, we consider a simplified situation 
($m_{u,d}=0$ and $m_s \neq 0$) and calculate the corrections to the GL 
potential from nonzero $m_s$.  In Sec.\ V, we introduce a parametrization of 
the gap and analyze possible phases near the super-normal phase boundary at 
finite temperature.  For these phases, residual symmetries, the Meissner 
masses of gluons, and the gapped and gapless quark modes are also 
investigated.  Section VI is devoted to summary and concluding remarks.   

\section{Ginzburg-Landau potential}

     In this section, as an introduction to later sections, we first study a 
toy model that captures the essential features of such a hierarchical 
structure of phase transitions as encountered in dense quark matter.  Validity
of the various approximations adopted later will be also clarified by the 
analysis of the toy model.  We then construct a general form of the GL
potential with the quark masses such that it satisfies symmetry constraints. 

\subsection{A toy model}
      
      Let us assume two real order parameters, $X$ and $Y$, and consider a 
GL potential of the following form:
\beq
\Omega &=& \alpha (X^2+Y^2) + \delta (X^2-Y^2) \nonumber \\
 & & + \beta_1(X^2+Y^2)^2 + \beta_2(X^4+Y^4).
\label{eq:toy-GL}
\eeq
This is the most general form of the potential up to the quartic order with 
the ``parity" symmetry,  $X\leftrightarrow -X$ and $Y \leftrightarrow -Y$.
For the stability of the potential, $\beta_1+\beta_2 >0$ and 
$2\beta_1+\beta_2>0$  are implicitly assumed.

      As we will see later, the coefficient $\alpha$ corresponds to 
the reduced temperature in the chiral limit, while $\delta$ denotes the 
degree of flavor symmetry breaking (in the present context, the breaking of 
$X^2 \leftrightarrow Y^2$ symmetry) which is proportional to the quark mass 
squared.  The coefficients $\beta_1$ and $\beta_2$ are the quartic couplings 
that do not change sign around the critical point, but more or less receive
corrections from the quark mass.  Bearing in mind the results that will be
obtained in later sections, we parametrize these coefficients as
 \beq
\label{eq:toy-alpha}
 \alpha &=& \mu^2 t = \mu^2 \frac{T-T_c}{T_c}, \\
\label{eq:toy-delta}
 \delta&=& \mu^2 \sigma = \mu^2 \frac{m^2}{g\mu^2} ,\\
\label{eq:toy-beta}
 \beta_{1}=\beta_2 &\equiv& \beta= \frac{\mu^2}{T_c^2},
 \eeq
where $\mu$ and $T_c$ are the quark chemical potential and the critical 
temperature in the chiral limit.  $m$ and $g$ are the quark mass and the QCD 
coupling constant, respectively.  In the last equalities of Eqs.\ 
(\ref{eq:toy-alpha})--(\ref{eq:toy-beta}), we mimic the actual 
parametric forms of $t$, $\sigma$, and $\beta$ shown in Eqs.\ (\ref{b1b2}) and 
(\ref{eq:full-sigma}) except for numerical factors.  We do not have to 
consider the quark mass corrections to $\beta_1$ and $\beta_2$ under the 
condition to be shown shortly.  $\delta$ is a key parameter which governs the 
multiple structure of the phase transitions: Due to the presence of $\delta$, 
the temperatures where the coefficients affixed to $X^2$ and $Y^2$ change sign 
are no longer identical.

      The minimum of the potential can be simply obtained from
$\partial \Omega/\partial X=\partial \Omega/\partial Y=0$.  Then one finds 
three different phases:
 \begin{enumerate}
 \item  a phase where both $X$ and $Y$ have non-vanishing
  condensates,
  \beq
\label{eq:XY-1}
  X^2&=& -\frac{1}{6\beta}(\alpha+3\delta)=
          - \frac{T_c^2}{6} (t+ 3 \sigma), \\
\label{eq:XY-2}
  Y^2&=& -\frac{1}{6\beta}(\alpha-3\delta)=
         - \frac{T_c^2}{6} (t- 3 \sigma). 
 \eeq
 \item a phase where only $Y$ has non-vanishing 
 condensate,
  \beq
  X^2&=& 0, \\
  Y^2&=& -\frac{1}{4\beta}(\alpha-\delta)=
         - \frac{T_c^2}{4} (t-   \sigma). 
 \eeq
\item a phase where there are no condensates,
 \beq
  X^2=Y^2=0.
 \eeq
\end{enumerate}

\begin{figure}[t]
\begin{center}
\includegraphics[scale=0.45]{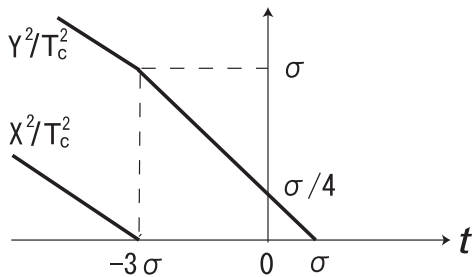}
\end{center}
\vspace{-0.5cm}
\caption{ Condensates $X^2$ and $Y^2$ as a function of the
  reduced temperature $t$ in a toy model. $\sigma$ is 
  the parameter that breaks ``flavor" symmetry $X^2 \rightarrow Y^2$.
}
\label{fig:toy-model}
\end{figure}

     The behavior of the condensates as a function of the reduced temperature
$t$ is illustrated in Fig.\ \ref{fig:toy-model}.  The multiple structure of 
the phase transitions induced by $\sigma$ shown in this figure is a general 
feature which survives in a realistic situation that will be discussed in later
sections.
   
     The questions to be addressed here are (i) what is the region of 
temperature where the GL potential Eq.\ (\ref{eq:toy-GL}) expanded up to the 
quartic order is valid, and (ii) what is the justification of neglecting the 
quark mass corrections to $\beta_1$ and $\beta_2$.  These questions can be 
answered if we restrict ourselves to the region
\beq   
 |t| < {\rm const} \cdot \sigma.
\label{eq:t-sig}
\eeq
Within this interval, both $X^2$ and $Y^2$ are of order or even smaller than
$\sigma T_c^2$.  Therefore, the dimensionless expansion parameters
of the GL potential satisfy
\beq
\frac{(X^2,Y^2)}{T_c^2}  < {\cal O}(\sigma),  
\label{eq:t-sig2}
\eeq  
which in turn guarantees that the higher order terms beyond the quartic order 
are negligible.  Furthermore, the additional quark mass corrections to 
$\beta_1$ and $\beta_2$ give only higher order corrections to $X^2$ and $Y^2$ 
as can be seen from, e.g., Eqs.\ (\ref{eq:XY-1}) and (\ref{eq:XY-2});
$X^2$ and $Y^2$ are already proportional to $\sigma$ if Eq.\ (\ref{eq:t-sig}) 
is satisfied.

\subsection{General form of GL potential}

     In this subsection, we construct a general form of the GL potential on 
the basis of the QCD symmetry under $G=SU(3)_C\times SU(3)_L\times SU(3)_R
\times U(1)_B$.
 
     For this purpose, let us first consider a basic variable, the gap 
function $\phi_{bcjk}$.  It is defined through the pairing gap $\Delta$ for a 
quark (color $b$ and flavor $j$) and another quark (color $c$ and flavor $k$) 
on the Fermi surface as will be shown in Sec.\ III [in particular, see 
Eqs.\ (\ref{eq:Del-phi}) and (\ref{G})].  Focusing our attention on 
spin-zero pairings of positive energy quarks in anti-symmetric combinations 
of colors and of flavors, we may rewrite $\phi_{bcjk}$ as \cite{IB} 
\begin{eqnarray}\label{phi-d}
[\phi_{bcjk}]_{L,R}=\epsilon_{abc} \epsilon_{ijk} [d_a^i]_{L,R} .
\end{eqnarray}
Here the 3$\times $ 3 matrix $[d_a^i]_{L}$ belongs to the 
{\rm fundamental} representation of $SU(3)_C  \times SU(3)_{L}$.  Similar 
properties hold for $[d_a^i]_{R}$ under the change $R \leftrightarrow L$.
In this subsection, we omit the color and flavor indices and write the gap 
function as ${\rm d}_{L,R}$ for simplicity.

     Under the operation of $G_{L,R}=SU(3)_C \times SU(3)_{L,R}\times U(1)_B$,
the left- and right-handed quarks and the gap ${\rm d}_{L,R}$ transform as
\begin{eqnarray}
\psi_{L,R}&\rightarrow& e^{i\varphi}U_C U_{L,R}\psi_{L,R}, \\
{\rm d}_{L,R}&\rightarrow& e^{-2i\varphi}U_{L,R}{\rm d}_{L,R}U_C^{\rm T},
\label{transform1}
\end{eqnarray}
where $U_{L,R}$'s are $3\times 3$ unitary matrices corresponding to each 
$SU(3)_{L,R}$ symmetry.  The phase factor is associated with the
$U(1)_B$ rotation.

     We now try to incorporate small but nonzero quark masses and write down
possible terms of the GL potential allowed by the QCD symmetry $G$.  In QCD 
with quark masses, we have such a term as 
\begin{equation}
{\cal L}=\bar{\psi}_L m \psi_R+h.c.,
\label{quarkmass}
\end{equation}
with $m$ being a $3\times 3$ mass matrix in flavor space.  Just like the 
standard procedure to construct the chiral Lagrangians, $m$ is assumed to 
transform in the following way \cite{Georgi},
\begin{equation}
m\rightarrow U_L m U_R^{\dagger}. 
\label{masstransform}
\end{equation}

     Following the transformation laws, Eqs.\ (\ref{transform1}) and
(\ref{masstransform}), we can write down a possible form of the GL potential 
as an expansion in terms of the order parameter, which is, in the
present context, the on-shell gap function ${\rm d}_{L,R}$.  Following the 
discussion in Sec.\ II.A, we take into account the quark mass terms up to 
${\cal O}(m^2)$ only in the quadratic terms in ${\rm d}_{L,R}$ and obtain
\begin{eqnarray}\label{generalGL}
\Omega &=& (a_0)_L {\rm Tr} ({\rm d}_L^{\dagger}{\rm d}_L)
+(a_0)_R {\rm Tr} ({\rm d}_R^{\dagger}{\rm d}_R) \\ \nonumber
& &+a_1[{\rm Tr}  ({\rm d}_L^{\dagger}m {\rm d}_R)+ h.c.] \\ \nonumber
& &+(a_2)_L {\rm Tr} ({\rm d}_L^{\dagger}mm^{\dagger}{\rm d}_L)
+(a_2)_R {\rm Tr}  ({\rm d}_R^{\dagger}m^{\dagger}m {\rm d}_R) \\ \nonumber
& &+(\hat{a}_2)_L{\rm Tr}  ({\rm d}_L^{\dagger} {\rm d}_L)  
{\rm Tr} (m^{\dagger}m ) +(\hat{a}_2)_R 
{\rm Tr}  ({\rm d}_R^{\dagger} {\rm d}_R)  {\rm Tr} (m^{\dagger}m ) \\ 
\nonumber
& & +c[{\rm det} m\cdot {\rm Tr}({\rm d}_R^{\dagger}m^{-1}{\rm d}_L)+ h.c.] \\
\nonumber
& &+(b_{1})_L [ {\rm Tr}  ({\rm d}_L^{\dagger}{\rm d}_L)]^2
+(b_{1})_R [{\rm Tr}  ({\rm d}_R^{\dagger}{\rm d}_R)]^2 \\ \nonumber
& &+ (b_{2})_L {\rm Tr}  [({\rm d}_L^{\dagger}{\rm d}_L)^2]
+(b_{2})_R {\rm Tr}  [({\rm d}_R^{\dagger}{\rm d}_R)^2] \\ \nonumber
& &+b_3 {\rm Tr} [({\rm d}_L^{\dagger}{\rm d}_L)
({\rm d}_R^{\dagger}{\rm d}_R)]
+b_4 {\rm Tr} ({\rm d}_L^{\dagger}{\rm d}_L) 
{\rm Tr} ({\rm d}_R^{\dagger}{\rm d}_R),
\end{eqnarray}
where the $a$'s, $b$'s, and $c$ are the expansion coefficients being
functions of temperature and chemical potential.

     Only the term proportional to $a_1$ in Eq.\ (\ref{generalGL}) breaks  
$U(1)_A$ symmetry and thus originates from anomaly induced interactions.
It does not affect the phase structure near the super-normal phase boundary 
at asymptotically high density as will be shown in Sec.\ IV.D.  The term 
proportional to $c$ also does not play an important role near the phase 
boundary as will be discussed at the end of Sec.\ III.B.

     We can further take into account the effects of electric and color 
neutralities and discuss possible terms to be included in 
Eq.\ (\ref{generalGL}).  In particular, the electric charge neutrality plays 
a central role in this paper.  Instead of writing down its general structure,  
we will show its explicit form calculated in the weak coupling regime in 
Sec.\ III.  The effect enters into the quadratic terms of the GL potential.
As for the color neutrality, its effect is subdominant in the vicinity of the 
phase boundary as has been already shown in Ref.\ \cite{IB}; the color 
neutrality affects only the quartic part of the GL potential (see Sec.\ IV.C).

\section{Corrections to the GL potential in weak coupling}
 
      In this section, we calculate corrections to the GL potential by 
unequal quark masses  ($m_{i}$, $i$=$u, d, s$) and unequal quark chemical 
potentials ($\mu_{i}$, $i$=$u, d, s$).  To simplify the derivation, we make 
several ansatze for the pairing gap:
\begin{enumerate}
\item 
The pairing between quarks is assumed to be in the zero total angular 
momentum ($J=0$) channel \cite{PR-PRD60}.  Furthermore, it is assumed to be
in the $LL$ and $RR$ channel as in the massless case and in the
positive parity channel as in the presence of $U(1)_A$ breaking \cite{RWA}.  
Then the gap can be written as 
$\Delta(k)=\gamma^5 \Delta^{(1)}+ {\bf \gamma} \cdot {\bf {\hat k}} 
\gamma^0 \gamma^5 \Delta^{(2)}$ \cite{BL}.

\item
The pairing is projected onto positive energy states.  In this case,
it is convenient to rewrite $\Delta(k)$ as  
\begin{eqnarray}
\Delta(k)=\gamma^5 \phi(k_0, {\bf k}) \Lambda^+ ({\bf k}),
\label{eq:Del-phi}
\end{eqnarray}
where
$\Lambda^+ ({\bf k})=(1+\gamma^0 {\bf \gamma} \cdot {\bf {\hat k}}) /2$
is a projection operator onto the positive energy states of 
massless quarks.

\item
The pairing is assumed to take place in the color antisymmetric and 
flavor antisymmetric channel.  This color channel is indeed the most 
attractive in weak coupling \cite{SON99,brown,PR}, while the flavor channel
is chosen in such a way as to satisfy the Pauli principle. 
Then, the pairing gap of a quark (color $b$ and flavor $j$) and another quark 
(color $c$ and flavor $k$) at the Fermi surface takes the form as given in 
Eq.\ (\ref{phi-d}).  Furthermore, the pairing in the positive parity channel
is represented by 
\begin{eqnarray}\label{parity-even}
{\rm d}_L={\rm d}_R\equiv d.
\end{eqnarray}

\item
We adopt the fact that the three-momentum dependence of the on-shell gap
$\Delta(q_0=\epsilon({\bf q}), {\bf q})$ or, equivalently, 
$\phi({\bf q};T)\equiv
 \phi(q_0=\epsilon({\bf q}), {\bf q})$
can be approximately factorized as \cite{PR}
 \begin{equation}
 \phi({\bf q};T) \simeq \phi(|{\bf q}|=\mu ; T) f({\bf q}),
 \label{approx1}
\end{equation}
where
 \begin{eqnarray}\label{f}
 f({\bf q})&=& \sin {{\bar g}y}|_{T=0}, \\
{\bar g}&=& g/ 3 \sqrt{2} \pi, \\
b &\equiv& 256 \pi^4 (2/3g^2)^{5/2}, \label{b} \\ 
 y({\bf q})&=&\ln \left[ \frac{2b \mu}{ ||{\bf q}|-\mu| + E({\bf q})} \right],
   \\
 E({\bf q})&=&\left[  ||{\bf q}|-\mu|^2+
 |\phi({\bf q};T)|^2 \right]^{1/2}. 
\end{eqnarray}

\item
We will assume later in Sec.\ V that $d_a^i$ is diagonal in color-flavor 
space even under the influence of nonzero quark masses and charge chemical 
potentials.  Then we will minimize the GL potential within the subspace.

\end{enumerate}

     In the following, we start with a known result for massless quarks 
(Sec.\ III.A) and then proceed to describe a more general situation with 
unequal quark masses  ($m_{i}$, $i$=$u, d, s$) and unequal quark chemical 
potentials ($\mu_{i}$, $i$=$u, d, s$) in Sec.\ III.B.
 
\subsection{Case with massless three flavors}
 
       For a homogeneous system composed of massless quarks $(m_{u,d,s}=0)$, 
quarks have a common chemical potential $\mu$, and the GL potential near the 
critical temperature $T_c$ expanded up to quartic order reads 
\cite{IB,PIS00}
\begin{eqnarray}
\label{GL}
   \Omega_0= \bar{\alpha} \sum_{a}|{\mathbf{d}_a}|^2
   +\beta_1(\sum_{a}|{\mathbf{d}_a}|^2)^2
   +\beta_2\sum_{ab}|{\mathbf{d}_a}^{\ast}
    \cdot {\mathbf{d}}_b|^2,  \nonumber \\
\end{eqnarray}
where $({\mathbf{d}_a})^i\equiv (d_a^u, d_a^d, d_a^s)$, and
the inner product is taken for flavor indices.   Using Eq.\ 
(\ref{parity-even}), one can relate the coefficients in Eq.\ (\ref{generalGL})
to those in Eq.\ (\ref{GL}) as
\begin{eqnarray}
&&\beta_1=(b_{1})_{L}+(b_{1})_{R}+b_4, \nonumber \\
&&\beta_2=(b_{2})_{L}+(b_{2})_{R}+b_3, \nonumber \\
&& \bar{\alpha}=(a_0)_L+(a_0)_R .
\end{eqnarray}
This potential is manifestly invariant under $SU(3)_C \times SU(3)_{L+R} 
\times U(1)_{B}$ rotation.  In the weak coupling approximation where the 
one-gluon exchange force in the normal medium is responsible for the pairing, 
the coefficients have been calculated as \cite{IB}
\begin{eqnarray}
\label{b1b2}
 \beta_{1}=\beta_{2}
  =\frac{7\zeta(3)}{8(\pi T_c)^{2}}N(\mu)
 \equiv \beta, \ 
  \bar{\alpha}=4 N(\mu) t \equiv \alpha_0  t. 
\end{eqnarray}
Here $N(\mu) = \mu^2/2\pi^2$ is the density of states at the Fermi surface, 
and $t=(T-T_c)/T_c$ is the reduced temperature.  With the parameters 
in Eq.\ (\ref{b1b2}), one finds a single second-order phase transition at 
$T=T_c$ from the CFL phase ($ {d}_a^i \propto \delta_a^i$) to the 
normal phase ($ {d}_a^i = 0$) in the mean-field theory \cite{IB}.  

\subsection{Case with unequal quark masses and chemical potentials}

     For a homogeneous system composed of unequal quark masses, there arise 
differences in the chemical potential among flavors and among colors
due to charge and color neutrality conditions.  The GL potential in this case 
is obtained in a similar approach to that adopted in Ref.\ \cite{IB} where a 
correction to Eq.\ (\ref{GL}) from color neutrality is calculated.

     First we start with the Nambu-Gor'kov two component field
($\psi_{ai}, \bar{\psi}_{ai}^{C}$), where $\psi^C=C \bar{\psi}^{\rm T}$ is
the charge-conjugate spinor.  The quark propagator of this field with 
the Hartree-Fock contributions ignored in the diagonal part reads
\begin{eqnarray}
 G(k) &\equiv&
\left(
\begin{array}{cc} 
   G^{(11)}(k)& G^{(12)}(k) \\
  G^{(21)}(k)& G^{(22)}(k) 
 \end{array}
\right) \\ 
& =&  \left(
\begin{array}{cc} 
  \gamma k+\gamma^{0}{\cal M} -m & {\tilde \Delta}(k) \\
  \Delta(k) & \gamma k-\gamma^{0}{\cal M}^{\rm T}-m 
  \end{array}
\right)^{-1} .
  \label{G}
\end{eqnarray}
Here ${\tilde \Delta}=\gamma^{0}\Delta^{\dagger}\gamma^{0}$, and 
\begin{eqnarray}
 {\cal M}_{abij}=\delta_{ab}\delta_{ij} \mu_{i} ,\ \ \ 
  m_{abij}=\delta_{ab}\delta_{ij}  m_{i}, 
\end{eqnarray}
are the quark chemical potentials and the quark masses in color ($a, b$) and 
flavor ($i, j$) space, respectively.  As in Sec.\ III.A, we again define 
$\mu$ as a quark chemical potential in the chiral limit ($m_{u,d,s}=0$).  
Once unequal quark masses are included under charge neutrality and $\beta$ 
equilibrium, $\mu_{i}$ differs from $\mu$.  We define the deviation of 
${\cal M}_{abij}$ from the massless case as
$\delta {\cal M}_{abij}\equiv \delta_{ab}\delta_{ij} (\mu_{i}-\mu) $.

     The free quark propagator $G_0$ and the self-energy $\Sigma$ are defined 
from the diagonal and off-diagonal components of $G^{-1}$ as usual:
\begin{eqnarray}
   G^{-1}(k)=G_{0}^{-1}(k) - \Sigma(k)\ .
 \label{GSig}
\end{eqnarray}

      Then the GL potential, which is a difference between the superfluid and 
normal phases near the critical temperature, is written in the CJT form 
\cite{CJT}
\begin{eqnarray}
&&\Omega = \Omega_{1} + \Omega_{2} \label{omegagl} \\
&&\Omega_{1} = \frac{T}{2}\sum_{n}
 \int\frac{d^{3}q}{(2\pi)^{3}}  
 {\rm Tr}[-G(q)\Sigma(q)+\ln G_{0}^{-1}(q)G(q)] , \nonumber \\
  \label{omega1} \\ 
&&\Omega_{2} = g^2 \frac{T^2}{4} 
 \sum_{m, n} 
\int\frac{d^{3}k}{(2\pi)^{3}} \int\frac{d^{3}q}{(2\pi)^{3}} \nonumber \\
&&~~  \! \! \times {\rm Tr}\Bigl[  
 {\cal D}_{\mu\nu}^{\alpha\beta} (q-k) 
\gamma^{\mu} \frac{\lambda^{\alpha}}{2} G^{(12)}(k)
\gamma^{\nu} {\left( \frac{\lambda^{\beta}}{2} \right)^{\rm T}} G^{(21)}(q) 
\nonumber \\
&&\! \! +{\cal D}_{\mu\nu}^{\alpha\beta} (q-k) 
 \gamma^{\mu} {\left(\frac{\lambda^{\alpha}}{2}\right)^{\rm T}}G^{(21)}(k)
\gamma^{\nu} \frac{\lambda^{\beta}}{2}G^{(12)}(q)
 \Bigl] . 
\label{omega2}
\end{eqnarray}
Here $\Omega_2$ represents the two particle irreducible graphs in the 
mean-field approximation with ${\cal D}(q)$ being the gluon propagator 
in hard dense loop approximation, and $\lambda^{\alpha}$ the generators of 
color $SU(3)$.  The summations are taken over Matsubara frequencies of quarks. 

     In the following, we will consider the corrections from 
$ \delta{\cal M}_{abij}$ and $m_{abij}$ to the quadratic term in $\Delta$ 
assuming that the corrections are small.  The corrections to the quartic terms 
are negligible near the critical temperature.  Figure 2 represents the
${\cal O}(\Delta ^2)$ contributions to $\Omega_{1}$ and $\Omega_{2}$. 

     First we consider $\Omega_{1}^{\Delta ^2}$,
\begin{eqnarray}
\Omega_{1}^{\Delta ^2}
 &=&-\frac{T}{2}\sum_{n~ {\rm odd}}
 \int\frac{d^{3}q}{(2\pi)^{3}}
  {\rm Tr} \Biggl[ \frac{1}{\gamma q+\gamma^{0}{\cal M}-m}
  {\tilde\Delta}(q) \nonumber\\
&&~~~~~~~~~\times \frac{1}{\gamma q-\gamma^{0}{\cal M}^{\rm T}-m}
\Delta(q)  \Biggl] \nonumber \\
&=& -\sum_{abij} \frac{i}{4} \oint \frac{d^{4}q}{(2\pi)^{4}}
 \tanh \left( \frac{q_0}{2T} \right)   {\rm Tr} \Biggl[   
 \frac{1}{\gamma q+\gamma^{0}{\mu}_{i}-m_{i}} \nonumber \\ &&~~~~~~~~  
\times {\tilde\Delta}(q)_{abij}    
\frac{1}{\gamma q-\gamma^{0}{\mu}_{j}-m_{j}}
\Delta(q)_{baji} \Biggl].
\label{eq:O-1-D}
\end{eqnarray}  
The first quark propagator in the right-hand side of Eq.\ (\ref{eq:O-1-D})
has particle and anti-particle poles,
\begin{eqnarray} \label{pp}
\epsilon^{p \pm}&=& \pm |{\bf q}| - p_F^{i} ,
\end{eqnarray}
while the second propagator has hole and anti-hole poles,
\begin{eqnarray} \label{hp}
\epsilon^{h \mp}&=& \mp |{\bf q}| + p_F^{j} ,
\end{eqnarray}
where $p_F^{i}=\sqrt{\mu_i^2-m_i^2}$ is the Fermi momentum of flavor $i$.
We define the shift of the chemical potential of flavor $i$ from $\mu$ as 
\begin{eqnarray}
\delta \mu_i = \mu_i -\mu.
\end{eqnarray}
Then, up to ${\cal O}(m^2/\mu^2, \delta \mu/\mu)$, $p_F^{i}$ can be written as 
\begin{eqnarray}\label{fermi-mom}
 p_F^{i} \simeq \mu_{i}-m_{i}^2/2\mu.
\end{eqnarray} 

     Of the poles in Eqs.\ (\ref{pp}) and (\ref{hp}), only two of them, 
$q_0= \epsilon^{p +}$ (particles) and $q_0= \epsilon^{h -}$ (particle holes),
have relevant contributions to the potential.  The other two poles contribute 
to the gaps on antiparticle (antiparticle hole) mass shell, which are 
irrelevant in our calculation.  Let us define 
\begin{eqnarray}
{\tilde q}&=&|{\bf q}|-\mu,  \\
{\cal F} ({\tilde q}) &=&\frac{\tanh ({\tilde q}/2T)}{{\tilde q}} .
\end{eqnarray}
Then, combining the residues of the two poles, we obtain
 \begin{eqnarray} \label{tree0}
 \Omega_{1}^{\Delta ^2}=
\sum_{abij}  \frac{1}{2}
\int \frac{d^3 q}{(2 \pi )^3}
\left( 
{\cal F} ({\tilde q}) 
-\frac{\partial {\cal F}  ({\tilde q}) 
 }{\partial{\tilde q}} \delta p_{ij} \right)  
 |\phi_{abij}|^2  .
\end{eqnarray}
Here, 
\begin{eqnarray}
 \delta p_{ij}  = (p_F^i + p_F^j ) /2 -\mu.
\end{eqnarray}
For later purpose, we also introduce an averaged shift of the chemical 
potential as
\begin{eqnarray}
 \delta \mu_{ij}  = (\delta \mu_i + \delta \mu_j )/2.
\end{eqnarray}

\begin{figure}[b]
\begin{center}
\includegraphics[width=8cm,height=4.5cm]{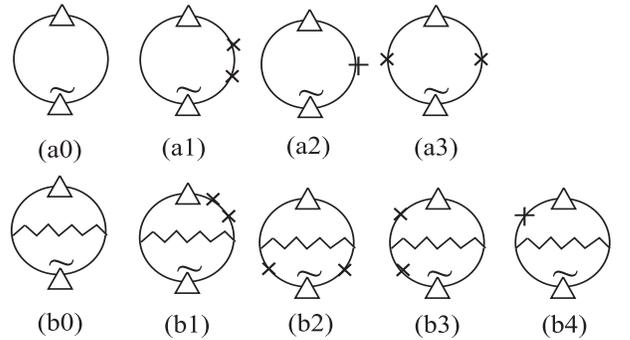}
\end{center}
\vspace{-0.5cm}
\caption{The quadratic terms of the GL potential.  $\times$ and $+$
represent $m$ and $\delta \mu$, respectively. (a0)--(a3) represent 
$\Omega_1^{\Delta^2}$, and (b0)--(b4) are the two particle irreducible graphs 
$\Omega_2^{\Delta^2}$.  (a0) and (b0) are the quadratic terms of the potential 
in the $m=\delta \mu=0$ case.  Among these graphs, (a3) and (b2) vanish 
due to the Lorentz structure of the gap.  (b3) vanishes because we consider 
the gaps constructed by quarks having the same chirality.  Only (a1), (a2), 
(b1), and (b4) give finite correction to the original GL potential 
[(a0) plus (b0)].}
\label{fig:graphs}
\end{figure}

     In the diagrammatic language, the first term in the right-hand side of 
Eq.\ (\ref{tree0}) corresponds to (a0) in Fig.\ \ref{fig:graphs}.  The second 
term corresponds to (a1) and (a2).  (a3) vanishes because of the Lorentz 
structure of the gaps together with the identity, 
${\rm Tr} (\Lambda^+ \Lambda^-) =0$.

     To perform the momentum integration in Eq.\ (\ref{tree0}), it is 
convenient to introduce a positive constant  $\kappa \sim {\cal O}(g^{-1})$ 
which cancels at the end of the calculation \cite{PR,IB}.  The cutoff of the 
three-momentum $\Lambda$ is taken to be $b\mu$ for the sake of consistency 
with the formula, Eq.\ (\ref{f}).  We divide the region of the momentum 
integral $-\mu < |{\bf q}|-\mu < b \mu $ into two parts, 
$0< ||{\bf q}|- \mu| < \kappa \phi_0$ (region I), and 
$-\mu < |{\bf q}|- \mu < -\kappa\phi_0$ 
and $\kappa\phi_0< |{\bf q}|- \mu< b\mu$ (region II) \cite{PR}.  Here we 
defined $\phi_0 \equiv [(1/9)\sum_{abij}|\phi_{abij}(\mu, T=0)|^2]^{1/2}$. 
In each region, the following approximations can be safely taken: in region I,
$\sin {{\bar g}y}|_{T=0} \sim 1$, and in region II, $\tanh 
\left[ (|{\bf q}| -\mu)/2T \right] \sim (|{\bf q}| -\mu) /||{\bf q}| -\mu|$.

     The first term in the right-hand side of Eq.\ (\ref{tree0}) is 
calculated as
 \begin{eqnarray} 
{\rm (a0)}&=& \sum_{abij}  \frac{1}{2}
\int \frac{d^3 q}{(2 \pi )^3}
 \ {\cal F} ({\tilde q}) \ 
 |\phi_{abij}|^2   \nonumber \\
&=&
\sum_{abij} 
\frac{1}{2}N(\mu) \Biggl\{
2 \int_{0}^{\kappa \phi_0} 
\ d \tilde{q} \ {\cal F} ({\tilde q}) 
\nonumber\\
&-&
\Biggl( \int_{\ln(b \mu/ \kappa \phi_0)}^{\ln b}
+\int_{\ln(b \mu/ \kappa \phi_0)}^{0} \Biggl) 
dy  f^2(y)
\Biggl\}
 |\phi_{abij}|^2  \nonumber \\
&=&
\sum_{abij} 
N(\mu)
\left\{ \ln \left(\frac{T_c}{T} \right)+ \frac{\pi}{4 {\bar g}} \right\}
 |\phi_{abij}|^2 .  \label{omega--1}
\end{eqnarray}
In the first equality, we replaced the momentum in the measure by the density 
of states at the Fermi surface.  In the second equality, we used the weak 
coupling approximation $g \ll 1$ and expanded $f(y)$ in terms of $g$.
We also used the explicit $g$ dependence of $T_c$ and $\phi_0$ in weak 
coupling \cite{brown,PR}, 
$\ln(T_c/\mu) \sim \ln(\phi_0/\mu) \sim -3\pi^2/\sqrt2 g$,
to obtain the final form.

     The second term in the right-hand side of Eq.\ (\ref{tree0}) is more 
involved:
\begin{eqnarray}
 {\rm (a1)}&+& {\rm (a2)} \nonumber\\
&=& -\frac{1}{2}
\int \frac{d^3 q}{(2 \pi )^3}
\frac{\partial {\cal F}  ( {\tilde q} )}{\partial {\tilde q} } \ \delta p \  
 |\phi|^2  \nonumber\\
&=&\frac{1}{4 \pi^2}
\int _{-\mu}^{b\mu} \ d{\tilde q} \ ({\tilde q}+\mu)^2 
\{ {\cal F}({\tilde q})-{\cal F}({\tilde q}+ \delta p ) \}
f^2({\tilde q}) \   |\phi|^2 \ \nonumber    \\
&\sim& 
\frac{1}{4 \pi^2}
\Biggl[
\left\{  \int_{-\mu}^{b\mu}
-\int_{-\mu+ \delta p }^{b\mu+ \delta p } \right\}
  \ d{\tilde q} \ {\cal F}({\tilde q}) f^2({\tilde q}) ({\tilde q}+\mu)^2 
\nonumber\\
&&~~~~~~~~~+2 \ \delta p \  
\int _{-\mu}^{b\mu} \ d{\tilde q} \  {\cal F}({\tilde q}) f^2({\tilde q}) 
({\tilde q}+\mu)
\Biggl]  \ |\phi|^2  \      \nonumber\\
&\sim& 
\frac{1}{4 \pi^2}
\Biggl[ 
- {\cal F}(b\mu) f^2(b\mu ) (b\mu+\mu)^2 \nonumber\\
&&~~+2 
\int _{-\mu}^{b\mu} \ d{\tilde q} \ {\cal F}({\tilde q}) f^2({\tilde q}) 
({\tilde q}+\mu)
\Biggl] \ \delta p \  |\phi|^2,  \
\label{inte}  \
\end{eqnarray}
where we abbreviated internal indices for simplicity.  $f(b\mu)$ appearing in 
the first term of the last bracket in Eq.\ (\ref{inte}) is 
${\cal O}(g(\phi^2/(b \mu)^2))$ and can be neglected at high densities. 
Then the second term leads to
\begin{eqnarray}
{\rm (a1)}&+&{\rm (a2)} \nonumber\\
&=&
\frac{1}{2 \pi^2} \mu\ \delta p \  
\Biggl\{
2 \int_{0}^{\kappa \phi_0} 
 \ d {\tilde q} \ {\cal F}({\tilde q}) \nonumber\\
&-&
\Biggl( \int_{\ln(b \mu/ \kappa \phi_0)}^{\ln b}
+\int_{\ln(b \mu/ \kappa \phi_0)}^{0}\Biggl) 
dy  f^2(y)
\Biggl\} |\phi|^2 \nonumber\\
&=&
-N(\mu)
\frac{\delta p}{\mu} \ln \left( \frac{T_c}{\mu} \right)
 \  |\phi|^2. \label{omega--2}
\end{eqnarray}
Here we again expanded $f(y)$ in terms of $g$ and used the explicit $g$ 
dependence of $T_c$ in weak coupling to obtain the final form. 

     Then the sum of Eqs.\ (\ref{omega--1}) and (\ref{omega--2}) becomes 
\begin{eqnarray} \label{tree}
\Omega_{1}^{\Delta^2}&=& 
{\rm (a0)}+{\rm (a1)}+{\rm (a2)} 
\nonumber\\
&=&\sum_{abij}  
N(\mu)
\Biggl[
\left\{ \ln \left(\frac{T_c}{T} \right)+ \frac{\pi}{4 {\bar g}} \right\} 
\nonumber\\
&-&\frac{\delta p_{ij}}{\mu} \ln \left( \frac{T_c}{\mu} \right)\  
\Biggl]  |\phi_{abij}|^2 . 
\end{eqnarray}

     Let us next evaluate $\Omega_2^{\Delta ^2}$.  In the diagrammatic 
description, this corresponds to (b0)--(b4) in Fig.\ \ref{fig:graphs}
in which the quark propagator is expanded in powers of masses and 
chemical potential differences.  Of these graphs, (b2) vanishes because of the
Lorentz structure of the gaps, and (b3) vanishes because the gaps are 
constructed by quarks having the same chirality.  As a result, only (b0), 
(b1), and (b4) have a finite value.  We expand these diagrams by 
$t \equiv |T-T_c|/T_c$ around $T_c$ up to ${\cal O}(t)$ for (b0) and 
${\cal O}(1)$ for (b1) and (b4). 

     We use a relation between $T_c$ and $g$ when $m_i$ and $\delta \mu_i$ 
in the denominator of the quark propagator are taken to be zero \cite{IB},
\begin{eqnarray}\label{Tcg}
&&f({\bf q})=
-g^2 \frac{1}{2}\oint \frac{d^4 k}{ (2 \pi)^4} \tanh 
\left( \frac{k_0}{2T_c}\right) {\rm Tr}\bigg[
{\cal D}_{\mu\nu}^{\alpha\beta} (q-k) 
\gamma^{\mu}  \nonumber \\
&&\times
{\left( \frac{\lambda^{\alpha}}{2} \right)^{\rm T}}
 (\gamma k-\gamma^{0}{\cal M})^{-1} 
(\gamma k+\gamma^{0}{\cal M})^{-1}    \gamma^{\nu}  
 {\left( \frac{\lambda^{\beta}}{2} \right)} f({\bf k}) \bigg].\nonumber\\
\end{eqnarray}
This equation is derived by taking $T \rightarrow T_c$ in the gap equation 
with the decomposition Eq.\ (\ref{approx1}).

     Using Eq.\ (\ref{Tcg}) which relates $T_c$ and $g$, 
we can evaluate (b0), (b1), and (b4) in Fig.\ \ref{fig:graphs}. We derive (b0)
as follows.  In leading order in $t$, (b0) becomes
\begin{eqnarray}
{\rm (b0)} & \rightarrow &
\frac{i}{4}  \oint\frac{d^{4}q}{(2\pi)^{4}} 
\tanh \left(\frac{q_0}{2T_c}\right)  \nonumber\\
&&\times {\rm Tr}
\Biggl[ 
\biggl(
-g^2 T_c
 \sum_{n~ {\rm odd}}
\int\frac{d^{3}k}{(2\pi)^{3}} 
{\cal D}_{\mu\nu}^{\alpha\beta} (q-k) 
\gamma^{\mu} {\left( \frac{\lambda^{\alpha}}{2} \right)^{\rm T}} \nonumber\\
&&~~~~~~~~~~~\times G^{(21)}(k) 
\gamma^{\nu} {\left( \frac{\lambda^{\beta}}{2} \right)}
\biggl)
 G^{(12)}(q)
\Biggl] \Bigg|_{T=T_c}  \nonumber\\
&=& 
\frac{i}{4}\oint \frac{d^4q}{(2\pi)^4} \tanh\left(\frac{q_0}{2T}\right)  
\nonumber\\
&&~~~~~~~~~~~
\times {\rm Tr}\left  [ f({\bf q}) \phi(T) G^{(12)}(q) \right]  
\Bigg|_{T=T_c}   \nonumber\\
&=& 
\frac{i}{4}\oint \frac{d^4q}{(2\pi)^4} \tanh\left(\frac{q_0}{2T}\right)
{\rm Tr} [ \Delta(q) G^{(12)}(q)] \Bigg|_{T=T_c}   \nonumber\\
&=&-\Omega_{1}^{\Delta^2}|_{T=T_c}.
\end{eqnarray}

     The ${\cal O}(t)$ term of (b0) becomes
\begin{eqnarray} 
&&{\rm (b0)} \rightarrow \nonumber\\
&& (T-T_c)\frac{i}{4} \frac{d}{dT} \biggl\{
\oint \frac{d^4k}{(2\pi)^4} \tanh\left(\frac{k_0}{2T}\right) 
 {\rm Tr} [G^{(21)}(k){\tilde\Delta(k)}] \nonumber\\
& &+
\oint \frac{d^4q}{(2\pi)^4} \tanh\left(\frac{q_0}{2T}\right)
{\rm Tr} [ \Delta(q) G^{(12)}(q)]
\biggl\} \Bigg|_{T=T_c}   
\nonumber\\ 
&&=2(T-T_c)\frac{d}{dT}(-\Omega_{1}^{\Delta ^2})|_{T=T_c}.
\end{eqnarray}

     (b1) and (b4) can be calculated in a similar way, and they become
\begin{eqnarray} 
{\rm (b1)}+{\rm (b4)}= N(\mu)\frac{2\delta p_{ij} }{\mu} 
\ln \left( \frac{T_c}{\mu} \right)
\  |\phi_{abij}|^2. 
\end{eqnarray}
Adding (b1) and (b4) to (b0), we finally obtain
\begin{eqnarray} 
\Omega_{2}^{\Delta ^2}&=&{\rm (b0)}+{\rm (b1)}+{\rm (b4)} \nonumber\\
&=&-\Omega_{1}^{\Delta^2}|_{T=T_c}
+2 (T-T_c)\frac{d}{dT}(-\Omega_{1}^{\Delta ^2})|_{T=T_c} \nonumber\\
&&~~~+ N(\mu)\frac{2\delta p_{ij} }{\mu} \ln \left( \frac{T_c}{\mu} \right)
\  |\phi_{abij}|^2 \nonumber\\
&=&\sum_{abij} 
N(\mu) \left\{
  -\frac{\pi}{4 \bar{g}}+ 2\ln \left( \frac{T}{T_c} \right)  
+\frac{2\delta p_{ij} }{\mu} \ln \left( \frac{T_c}{\mu} \right)
 \ \right\} \nonumber\\
&&~~~~~~~~~~~~~~~~~~~~~~
\times  |\phi_{abij}|^2  .\label{correction}
\end{eqnarray}

     Combining Eqs.\ (\ref{tree}) and (\ref{correction}), we obtain the final 
result for the quadratic term of the GL free energy,
\begin{eqnarray}
 \Omega_{GL}^{\Delta ^2}
&=&\sum_{abij} N(\mu)
 \left\{
\ln \left( \frac{T}{T_c} \right)
-  \frac{\delta p_{ij} }{\mu} \ln \left( \frac{\mu}{T_c} \right) \right\}
 \nonumber \\
&&~~~~~~~~~~~~~~~~~~~~~~~~~~~~~~~~~~~~
\times |\phi_{abij}|^2, 
\label{eq:final-form-1} 
 \\
\delta p_{ij} &\simeq &
\delta \mu_{ij} - (m_i^2+m_j^2)/(4\mu).
\label{eq:final-form-2} 
\end{eqnarray}

     It can be rewritten in a clearer form,
\begin{eqnarray}
 \Omega_{GL}^{\Delta ^2}
&=&\sum_{abij} N(\mu)
\left( \frac{T- {T}_{c}^{ij}} {{T}_{c}} \right)| \phi_{abij}|^2, 
\label{eq:final-form-3} \\
\frac{{T}_{c}^{ij}}{T_c}&=&
1+  \frac{\delta p_{ij} }{\mu} 
\ln \left( \frac{\mu}{T_c} \right).
\label{eq:final-form-4}
\end{eqnarray}

     Several comments are in order here.
\begin{itemize}

\item The free energy correction depends only on the average shift of the 
Fermi momenta of paired quarks, $\delta p_{ij}$, as can be seen from 
Eq.\ (\ref{eq:final-form-1}).  Furthermore, the correction may be decomposed
into a sum of the effects from the averaged chemical potential and the 
averaged mass squared when $\delta p_{ij}$ is small, as shown in 
Eq.\ (\ref{eq:final-form-2}).
  
\item
Equations (\ref{eq:final-form-3}) and (\ref{eq:final-form-4}) indicate that 
the shift $\delta p_{ij}$ affects the free energy correction through the 
critical temperature $T_c^{ij}$ for the $(i,j)$ pairing.  The larger the 
shift $\delta p_{ij}$ is, the higher the melting temperature ${T}_{c}^{ij}$ 
becomes.  In other words, one can compare the thermal stability of two 
different pairs only by taking note of a difference in the average density 
of states between them.

\item
The unlocking at $T=0$ is different in mechanism from our unlocking near 
$T_c$.  At $T=0$, color-flavor unlocking due to the Fermi momentum 
{\em mismatch} between paired quarks is expected at 
$\mu \sim m_s^2 / \phi_0$ \cite{unlock}.  In contrast, the Fermi momentum 
{\em average} is important for our unlocking near $T_c$.

\item
We can see that in the correction obtained above quark flavors do not mix
with each other.  As long as we consider the pairing of positive energy quarks
and neglect that of antiquarks, the flavor structure of a gap always enters in 
the form of $m_i^2$ instead of $m_i m_j (i \neq j)$.
\end{itemize}

\section{GL potential in realistic quark matter}

      Up to now we constructed the GL potential for general quark masses 
and chemical potential shifts assuming that the corrections are small.
In this section, we will focus on the case close to a realistic situation 
by taking $m_{s} \neq 0$ with $m_{u,d}=0$ and by imposing $\beta$ 
equilibrium and charge neutrality conditions.  This analysis clarifies the 
role of the strange quark mass in possible phase structures near the 
super-normal phase boundary at finite $T$. 

\subsection{Direct $m_s$ correction through the quark mass term}

     Let us first consider the correction directly proportional to
$m_s^2$ in Eqs.\ (\ref{eq:final-form-1}) and (\ref{eq:final-form-2}).
It is easy to see from the flavor structure of Eq.\ (\ref{eq:final-form-1}) 
that $m_s$ affects only $us$ and $ds$ pairings.  This is also reasonable
from the physical point of view as will be shown below.

     By inserting $m_s\neq 0$ with $m_{u,d}=0$ and using Eq.\ (\ref{phi-d}) 
with Eq.\ (\ref{parity-even}), the direct $m_s$ correction in Eq.\ 
(\ref{eq:final-form-1}) becomes
\begin{eqnarray}\label{epsilon}
\epsilon 
\sum_{a} ( |d_a^u|^2 + |d_a^d|^2 )
= \epsilon \sum_{a} (|{\mathbf{d}_a}|^2 -|d_a^s|^2).   
\end{eqnarray}
Here the coefficient $\epsilon$ reads
 \begin{eqnarray}
\label{epsilon-wc}
\epsilon  \simeq   
\alpha_0 \frac{m_s^2}{4 \mu^2}
\ln \left(\frac{\mu}{T_c}\right)
\sim 2 \alpha_0  \sigma,
\end{eqnarray}
where $\alpha_0=4N(\mu)$ is defined in Eq.\ (\ref{b1b2}), and  
a dimensionless parameter $\sigma$ is introduced as
\begin{eqnarray}
\label{eq:full-sigma}
\sigma = \left( \frac{3 \pi^2} {8 {\sqrt 2}} \right) \frac{m_s^2} {g \mu^2}.
\end{eqnarray} 
In Eq.\ (\ref{epsilon-wc}) we have used a weak coupling relation, 
$\ln (T_c/\mu) \sim -{3\pi^2}/(\sqrt{2} g)$, which originates from the 
long-range color magnetic interaction \cite{brown,PR}. 
 
     Since the finite $m_s$ decreases the Fermi momentum of $s$ quarks,
$\delta p_{us}$ and $\delta p_{ds}$ are smaller than $\delta p_{ud}$, and
thus $\epsilon$ becomes positive such that $ud$ pairing is favored over $us$ 
and $ds$ pairings.  Consequently, the CFL phase becomes asymmetric in flavor 
space and its critical temperature is lowered, leading to the appearance of 
the 2SC phase (${d}_a^i \propto \delta^{is}$) just below $T_c$ 
\cite{abuki}. 

\subsection{Indirect $m_s$ correction through the charge chemical potentials}

      We proceed to discuss the correction from the charge chemical 
potentials, which is proportional to $\delta \mu_{ij}$ in Eqs.\ 
(\ref{eq:final-form-1}) and (\ref{eq:final-form-2}).  Under $\beta$ 
equilibrium and charge neutrality, the electron chemical potential $\mu_e$ and
the shift $\delta\mu_i$ of the chemical potential of flavor $i$ are related as
\begin{eqnarray}\label{delmu}
\delta \mu_i =-q_i\mu_e 
\end{eqnarray}
with $q_i$ being the electric charges of the quarks; $q_{u}=2/3$ and 
$q_{d,s}=-1/3$.  In weak coupling, where one may regard normal quark matter 
and electrons as noninteracting Fermi gases, $\mu_e$ is related to $m_s$ as
\begin{eqnarray}\label{muemus}
\mu_e=m_s^2/4\mu .
\end{eqnarray}
This estimate is valid in the vicinity of $T_c$ where corrections to $\mu_e$ 
by a finite pairing gap affect only the quartic terms in the GL potential.
 
     By inserting Eqs.\ (\ref{delmu}) and (\ref{muemus}) and using 
Eq.\ (\ref{phi-d}) with Eq.\ (\ref{parity-even}), the indirect $m_s$ 
correction in Eq.\ (\ref{eq:final-form-1}) becomes
\begin{eqnarray}\label{eta}
\eta~ (\frac{1}{3}  \sum_{a}|{\mathbf{d}_a}|^2
 -\sum_{a} |d_a^u|^2).  
\end{eqnarray}
Here the coefficient $\eta$ reads
 \begin{eqnarray}
 \label{eta-wc}
\eta \simeq
 \alpha_0 \frac{m_s^2}{8\mu^2}
\ln \left(\frac{\mu}{T_c}\right)
\sim \alpha_0 \sigma  .
\end{eqnarray} 
Since $\delta \mu_{ds}$ is larger than $\delta \mu_{ud}$ and $\delta \mu_{us}$
and it increases the average Fermi momentum of $ds$ quarks in Eq.\ 
(\ref{eq:final-form-2}), the indirect $m_s$ correction favors $ds$ pairing 
than $ud$ and $us$ pairings.

     Note that we only consider the modification of $T_c$ due to nonzero
$m_s$ through the properties of the superfluid phase.  Actually, $T_c$ is 
modified not only by the properties of the superfluid phase but also of the 
normal phase.  The modification by the normal phase enters through the Debye 
mass in the gluon propagator ${\cal D}$ in the relation between $T_c$ and $g$,
Eq.\ (\ref{Tcg}).  Fortunately the modification due to nonzero $m_s$ and 
charge neutrality in normal quark matter is like $T_c \rightarrow 
T_c (1+ {\cal O}(g \sigma))$.  This modification is of higher order than 
that in superfluid quark matter, $T_c \rightarrow T_c (1+ {\cal O}(\sigma))$, 
to be derived in Sec.\ V. 

\subsection{Effect of color neutrality}

    We consider color neutrality of the system as well.  In contrast to the 
case at $T=0$, however, it affects only the quartic terms in the GL potential 
because the possible chemical potential differences between colors are 
\cite{IB,RIS}
\begin{eqnarray}
    \delta{\mu}_{ab}&=&
  -\frac{1}{9\mu}\ln\left(\frac{T_{c}}{\mu}\right)
   [3\sum_i (d_{a}^i  d_{b}^{i*}) 
-  \delta_{ab} \sum_{c,i} |d_c^i|^{2}], \nonumber \\
  \label{muaGL}  
\end{eqnarray}
which are already of quadratic order in the gap.  Here $\delta \mu_{ab}= 
\mu_{ab} -\mu \delta_{ab}$, and  ${\mu}_{ab}$ is the chemical potential 
conjugate to $n_{ab}=\sum_i \bar{\psi}_{bi} \gamma^0 \psi_{ai}$ \cite{IB}.
In weak coupling the magnitude of the correction to the quartic terms is 
suppressed by ${\cal O}((T_c/g\mu)^2)$ compared to the leading quartic terms.
Thus color neutrality has no essential consequence to the phase transitions 
considered in this paper.  A major difference between the corrections from the 
charge neutrality and the color neutrality is that the former shifts the 
quark chemical potentials even in the normal phase, while the latter works 
only when the pairing occurs.  This is why the former is more important 
than the latter near $T_c$.
  
\subsection{Effect of instantons}
   
    In QCD with three flavors, instantons induce chirality-flipping
six-fermion interactions.  This interaction leads to a sextic term in the gap
in the chiral limit, which is irrelevant near the super-normal phase boundary.
If the strange quark mass enters, the direct instantons induce an effective 
four-fermion interaction between $u$ and $d$ quarks \cite{Sch}.  This leads to 
a quadratic term in the GL potential, $\xi~\sum_a |d_a^s|^2$, which 
corresponds to the $a_1$ term in Eq.\ (\ref{generalGL}).  An explicit 
calculation in weak coupling shows that $\xi\sim -\alpha_0 (m_s/\mu) 
(\Lambda_{\rm QCD}/\mu)^9 (1/g)^{14}$. 
The negative sign indicates that the instanton effect favors $ud$ pairing 
as does one-gluon exchange [see Eq.\ (\ref{epsilon})].  However, the magnitude
of $\xi$ is highly suppressed at high densities.  Therefore we will ignore 
this term in the following.

     In summary of Secs.\ IV.A--D, as far as we are close to the super-normal 
phase boundary at high density, we have only to consider the corrections
from the strange quark mass and the electric charge neutrality, which favor 
$ud$ pairing and $ds$ pairing, respectively. 

\subsection{Validity of the approximations}

     Our analysis of the GL potential near $T_c$ is valid as long as 
(a) $g \ll 1$, (b) $\sigma \ll 1$, and (c) $T_c \ll g \mu $.  The condition 
(a) is necessary for the dominance of the long-range magnetic interaction 
acting between quarks.  The condition (b) allowed us to consider the 
corrections to the GL potential by the strange quark mass and differences 
in the chemical potential among flavors up to quadratic order in the gap.
The condition (c) is imposed so that the correction to the quartic terms in 
the GL potential from the color neutrality can be neglected.  These conditions
are all satisfied at asymptotically high density.
  
     If we use the result for $T_c$ calculated in weak coupling 
\cite{brown,PR}, one finds that (c) is a consequence of (a).  The conditions 
(a) and (b) can also be combined into  
\begin{eqnarray}
\frac{m_s^2}{\mu^2} \ll g \ll 1.
\end{eqnarray}

\section{Melting pattern of diquark condensates}

     In this section, we clarify the phase structure near the transition
temperature using the GL potential corrected by Eqs.\ (\ref{epsilon}) and 
(\ref{eta}).  We start from assuming the form of the condensation and derive 
the temperature dependence of the gaps that minimize the potential.  Then we 
discuss the properties of quark quasiparticles and transverse gluons in
each of the phases near the transition temperature.

\subsection{Phase structure with finite $m_s$ and charge neutrality}

     Since the two effects of nonzero $m_s$, characterized by Eqs.\ 
(\ref{epsilon}) and (\ref{eta}), favor $ud$ pairing and $ds$ pairing, 
respectively, the finite temperature transition from the CFL to 
the normal phase at $m_s=0$ is significantly modified.  In fact, as we show 
in detail below, successive color-flavor unlockings take place instead of a 
simultaneous unlocking of all color-flavor combinations.  To describe this 
{\em hierarchical thermal unlocking}, it is convenient to  
introduce a parameterization
\begin{eqnarray}
\label{s}
d_a^i
=
 \left(
  \begin{array}{lll} 
      \Delta_1&0&0    \\
  0&\Delta_2&0    \\
  0&0&\Delta_3    \\
  \end{array}
 \right).
  \end{eqnarray}
We assume $\Delta_{1,2,3}$ to be real.  We also name the phases for later 
convenience as \cite{IMTH04}
 \begin{eqnarray}
\label{phase-def} 
  \begin{array}{llcl} 
     \Delta_{1,2,3} \neq 0 & & :  &  {\rm mCFL}, \\
\Delta_1=0, & \Delta_{2,3}\neq 0   & :& {\rm uSC},      \\
\Delta_2=0, &  \Delta_{1,3} \neq 0 & :& {\rm dSC},      \\
\Delta_3=0, &  \Delta_{1,2} \neq 0 & :& {\rm sSC},      \\
\Delta_{1,2}=0, & \Delta_3 \neq 0  &: & {\rm 2SC},                
  \end{array}
  \end{eqnarray} 
where dSC (uSC, sSC) stands for superconductivity in which for $d$ ($u$, $s$)
quarks all three colors are involved in the pairing.
   
     In terms of the parameterization (\ref{s}), the GL potential with 
corrections of ${\cal O}(m_s^2)$ to the quadratic term, 
Eqs.\ (\ref{epsilon}) and (\ref{eta}), reads
\begin{eqnarray}
 \label{new-GL}
\Omega &=&\bar{\alpha}' (\Delta_1^2+\Delta_2^2+\Delta_3^2) 
- \epsilon \Delta_3^2 - \eta \Delta_1^2\nonumber \\
 &+& \beta_1 (\Delta_1^2+\Delta_2^2+\Delta_3^2)^2
 + \beta_2 ( \Delta_1^4+\Delta_2^4+\Delta_3^4),
\end{eqnarray}
where $\bar{\alpha}' =\bar{\alpha}+\epsilon+\frac{\eta}{3}$.

     We proceed to analyze the phase structure dictated by Eq.\ 
(\ref{new-GL}) with the weak coupling parameters (\ref{b1b2}), 
(\ref{epsilon-wc}), and (\ref{eta-wc}) up to leading order in $g$.
A stable condensation must satisfy the gap equations
\begin{eqnarray} 
\partial \Omega /\partial\Delta_{1,2,3}=0,
\end{eqnarray}
and it must minimize the potential Eq.\ (\ref{new-GL}).  We compare the 
energies of all condensates in Eq.\ (\ref{phase-def}) and decide the most 
stable condensation at each temperature near $T_c$.

     In Figs.\ 3 and 4 the results thus obtained for the phase structure 
near $T_c$ are summarized.

     Figure 3(a) shows the second-order phase transition, CFL $\to$
normal for $m_s=0$.  Figures 3(b,c) represent how the phase transitions and 
their critical temperatures bifurcate as we introduce (b) effects of a nonzero 
$m_s$ in the quark propagator and then (c) effects of charge neutrality.
In case (b), two second-order phase transitions arise:
mCFL (with $\Delta_1=\Delta_2$) $\to$ 2SC at 
$T=T_c^s \equiv (1-4 \sigma)T_c$, 
and 2SC $\to$ normal at $T=T_c$.  In case (c), there arise three 
successive second-order phase transitions, mCFL $\to$ dSC at $T=T_c^{\rm I}$,
dSC $\to$ 2SC at $T=T_c^{\rm II}$, and 2SC $\to$ normal at $T=T_c^{\rm III}$. 
Shown in Fig.\ 4 is the $T$-dependence of the gaps $\Delta_{1,2,3}$ for 
the case (c).  All the gaps are continuous functions of $T$, but their slopes
are discontinuous at the critical points, which reflects the second order 
nature of the transitions in the mean-field treatment of Eq.\ (\ref{new-GL}).

\begin{figure}[t]
\begin{center}
\includegraphics[width=7cm]{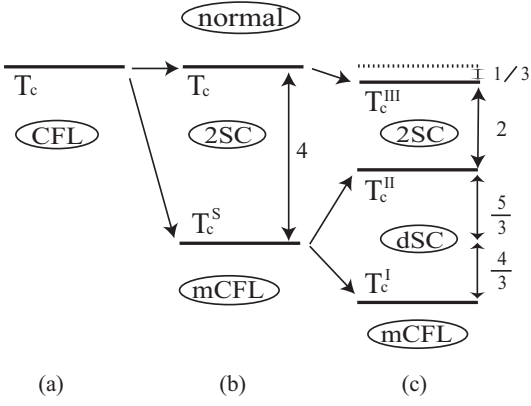}
\end{center}
\vspace{-0.5cm}
\caption{Transition temperatures of the 
three-flavor color superconductor in weak coupling:
(a) all quarks are massless;
(b) nonzero $m_s$ in the quark propagator is considered;
(c) electric charge neutrality is further imposed.
The numbers attached to the arrows are in units of $\sigma T_c$.  From Ref.\ 
\cite{IMTH04}.
}
\label{fig:Tc}
\end{figure}
     
\begin{figure}[t]
\begin{center}
\includegraphics[width=7cm]{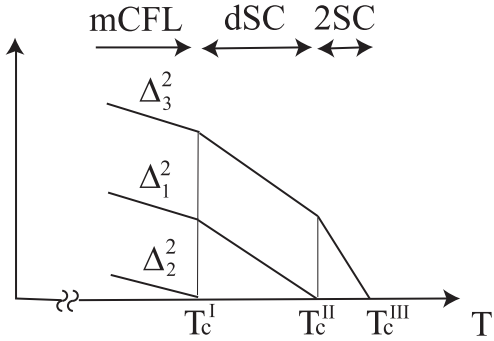}
\end{center}
\vspace{-0.5cm}
\caption{A schematic illustration of the gaps squared
as a function of $T$.  From Ref.\ \cite{IMTH04}.}
\label{fig:conden}
\end{figure}
  
     We may understand the bifurcation of the transition temperatures 
as follows.  In the massless case (a), $T_c$ is degenerate 
between the CFL and 2SC phases, the chemical potential is common to all
three flavors and colors, and the CFL phase is more favorable than
the 2SC phase below $T_c$.  As one goes from (a) to (b), the density of 
states of the $s$ quarks at the Fermi surface is reduced.  Then the critical 
temperature for the CFL phase is lowered, and the 2SC phase is allowed 
to appear at temperatures between $T_c^s$ and $T_c$.  As one goes from (b) 
to (c), the average chemical potential of $ds$ ($ud$ and $us$) quarks 
increases (decreases).  Accordingly, the transition temperatures
further change from $T_c$ to $T_c^{\rm III}$ and from $T_c^s$ to 
$T_c^{\rm I}$ and $T_c^{\rm II}$. 

     Now we examine in more detail how the color-flavor unlocking in case 
(c) proceeds with increasing $T$ from the region below $T_c^{\rm I}$.

 \noindent
 (i)
     Just below $T_c^{\rm I}$, we have a CFL-like phase, but the three gaps 
take different values, with an order $\Delta_3 > \Delta_1 > \Delta_2 \neq 0$ 
(the mCFL phase).  The reason why this order is realized can be understood 
from the GL potential (\ref{new-GL}).  The $\epsilon$-term and $\eta$-term in 
Eq.\ (\ref{new-GL}) tend to destabilize $us$ pairing ($\Delta_2$)
relative to $ud$ pairing ($\Delta_3$) and $ds$ pairing ($\Delta_1$), and
since $\epsilon > \eta (> 0)$, $ds$ pairing is destabilized more effectively 
than $ud$ pairing.  The value of each gap in the mCFL phase reads
    \begin{eqnarray} \label{gap1}     
\Delta_i^2=\frac{\alpha_0}{8\beta} \left( 
 \frac{T_c-T}{T_c} + c_i \sigma 
\right) ,  
\end{eqnarray}
with $c_{1,2,3}=(-4/3, -16/3, 8/3)$.
The mCFL phase has only a global symmetry $U(1)_{C+L+R} \times U(1)_{C+L+R}$ 
in contrast to the global symmetry $SU(3)_{C+L+R}$ in the CFL phase with 
$m_{u,d,s}=0$.  The remaining $U(1)$ generators are 
\begin{eqnarray}
{\cal Q}_1&=&T_3- S_3,  \nonumber\\
{\cal Q}_2&=&T_8-S_8.   
\end{eqnarray} 
Here $T_{a}$ ($S_{i}$) are the part of the generators of $SU(3)_C$ 
($SU(3)_{L+R}$) with explicit forms $T_3=S_3=({1}/{2}){\rm diag}(1,-1,0)$ and 
$T_8=S_8=({1}/{2 {\sqrt 3}}){\rm diag}(1,1,-2)$.  The free energy in this 
phase is
\begin{eqnarray}
\Omega_{mCFL}=-\frac{3\alpha_0^2}{16\beta}\left\{ \left(\frac{T-T_c}{T_c}
+\frac{4}{3}\eta \right)^2 
+\frac{8}{3}\sigma^2 \right\}.
\end{eqnarray}
As $T$ increases, the first unlocking transition, the unlocking of 
$\Delta_2$ (the pairing between $Bu$ and $Rs$ quarks), takes place at the 
critical temperature,
 \begin{eqnarray}
 \label{Tc-1}
T_c^{\rm I}  = \left( 1 - \frac{16}{3} \sigma \right) T_c  .
\end{eqnarray}

\noindent
(ii) For $T_c^{\rm I} < T < T_c^{\rm II}$, $\Delta_2=0$ and
\begin{eqnarray} \label{gap2}
\Delta_i^2= \frac{\alpha_0}{6 \beta} 
\left(
 \frac{T_c-T}{T_c} + c_i \sigma
\right) ,  
\end{eqnarray}
with $c_{1,3}=(-7/3, 2/3)$.
In this phase, we have only $ud$ and $ds$ pairings (the dSC phase), and there 
is a manifest symmetry, $U(1)_{C+L+R} \times U(1)_{C+L+R} \times 
U(1)_{C+V+B} \times U(1)_{C+V+B}$, where the corresponding $U(1)$ generators 
are 
\begin{eqnarray}
{\cal Q}_1&=&T_3- S_3, \nonumber \\
{\cal Q}_2&=&T_8-S_8,  \nonumber \\
{\cal Q}_3&=&Q+\frac{2}{\sqrt 3} T_8- 2 S_3, \nonumber \\
{\cal Q}_4&=&Q+\frac{2}{\sqrt 3} S_8- 2 S_3,  \label{qqq}
\end{eqnarray} 
respectively.
Here $Q={2}/{3}$ is the generator of baryon charge ($Q$) and acts on the gap 
as Eq.\ (\ref{transform1}).
The free energy in this phase is
\begin{eqnarray}
\Omega_{dSC}=-\frac{\alpha_0^2}{6\beta}\left\{\left(\frac{T-T_c}{T_c}+\frac{5}{6}\eta \right)^2 
+\frac{3}{4}\sigma^2 \right\}.
\end{eqnarray}
At $T=T_c^{\rm II}$, the second unlocking transition, the unlocking of 
$\Delta_1$ (the pairing between $Gs$ and $Bd$ quarks), takes place at the 
critical temperature,
   \begin{eqnarray}
 \label{Tc-2}
T_c^{\rm II}  = \left( 1 -\frac{7}{3}\sigma \right) T_c.
\end{eqnarray} 

\noindent
(iii) For $T_c^{\rm II} < T < T_c^{\rm III}$, one finds the 2SC phase, which 
has only $ud$ pairing with 
\begin{eqnarray} \label{gap3}
\Delta_3^2=\frac{\alpha_0}{4 \beta} 
\left(
 \frac{T_c-T}{T_c} 
-\frac{1}{3}\sigma
\right) .  
\end{eqnarray} 
The 2SC phase has a symmetry $SU(2)_C \times SU(2)_{L+R} \times U(1)_{C+B} 
\times U(1)_{L+R+B}$, where the corresponding $U(1)$ generators are
\begin{eqnarray}
{\cal Q}_5&=&Q+{\sqrt 3} T_8, \nonumber \\
{\cal Q}_6&=&Q+{\sqrt 3} S_8,  
\end{eqnarray} 
respectively.
The free energy in this phase is
\begin{eqnarray}
\Omega_{2SC}=-\frac{\alpha_0^2}{8\beta}\left(\frac{T-T_c}{T_c}+\frac{1}{3}\sigma\right)^2 .
\end{eqnarray}
The final unlocking transition where 
 $\Delta_3$ (the pairing between $Rd$ and $Gu$ quarks) vanishes occurs at 
\begin{eqnarray}
 \label{Tc-3}
T_c^{\rm III}  = \left( 1 -\frac{1}{3}\sigma \right) T_c .
\end{eqnarray}  
Above $T_c^{\rm III}$, the system is in the normal phase.  Note that in the 
temperature region $(T_c^{\rm I}-T)/T_c^{\rm I} < {\rm const} \cdot \sigma$, 
the gaps satisfy the condition ($\ref{eq:t-sig2}$), so the GL potential 
expanded up to quartic order in the gap is valid. 

\begin{figure}[t]
\begin{center}
\includegraphics[width=7cm]{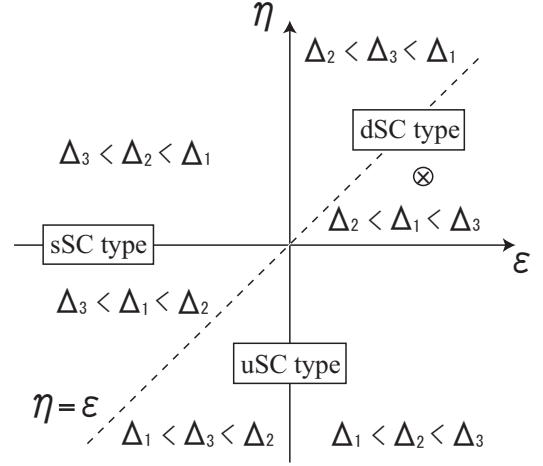}
\end{center}
\vspace{-0.5cm}
\caption{
The phase structure obtained from Eq.\ (\ref{new-GL}) in ($\epsilon, \eta$) 
plane for $\beta_1=\beta_2>0$.  The mCFL phase is assumed to appear at 
sufficiently low temperature.  $\otimes$ indicates the point calculated  
in the weak coupling approximation.
 }
\label{fig:epsilon-eta}
\end{figure}

     So far, we have derived the parameters $\beta_{1,2}$, $\epsilon$, and 
$\eta$ in the weak coupling approximation and discussed the phase structure 
expected at asymptotically high density.  Now we relax the weak coupling 
constraint and study possible phase structures expected from the GL potential 
of the form Eq.\ (\ref{new-GL}) with arbitrary coupling strengths.  To 
simplify the argument, we assume $\beta_1=\beta_2 >0$ and draw the phase 
diagram in the space of the parameters $\epsilon$ and $\eta$ in Fig.\ 
\ref{fig:epsilon-eta}.
  
     If we consider the case where the mCFL phase appears at temperature 
sufficiently below $T_c$, the relative magnitudes of the gaps behave
as shown in Fig.\ \ref{fig:epsilon-eta}.  If $\Delta_{2}$ is the smallest of 
the three gaps, we call the melting pattern ``dSC-type'' because the dSC phase 
would appear as $T$ increases.  ``uSC-type'' and  ``sSC-type'' are defined in 
a similar way.   As in Fig.\ \ref{fig:epsilon-eta}, depending on the relative 
magnitude of $\epsilon$ and $\eta$, the dSC-type melting pattern is divided
into two different classes in which the 2SC-type phase
($\Delta_1 \neq 0, \Delta_{2,3}=0$ or $\Delta_3 \neq 0, \Delta_{1,2}=0$) 
appears just below the transition temperature to the normal phase.
The uSC-type and sSC-type melting patterns also have such a two-fold
structure.  The {\em hierarchical thermal unlocking} (mCFL $\rightarrow$ 
  dSC  $\rightarrow$ 2SC  $\rightarrow$ normal) is realized in weak coupling 
as shown by $\otimes$ in the figure.  The figure indicates that 
it is a rather robust phenomenon near the super-normal phase boundary.
In other words, the uSC or sSC phase appears only when $\epsilon$ or
$\eta$ changes sign at least under the form of the GL potential in Eq.\ 
(\ref{new-GL}) with an assumption $\beta_1=\beta_2>0$.  We mention here
that a recent analysis in Ref.\ \cite{Fukushima} using the Nambu-Jona-Lasinio 
(NJL) model shows that the signs and the ratios of the coupling strengths in 
the GL potential take similar values with our weak coupling values:
$\beta_1/\beta_2=1$ (with $\beta_1>0$ and $\beta_2>0$)
and $\zeta \equiv \epsilon / \eta \cong  2$--4 (with $\epsilon >0$ and
$\eta >0$). 

 \subsection{Properties of quark and gluon modes}

     Next we resume considering about the weak coupling limit and discuss the 
properties of quark and gluon modes such as the excitation energies 
in each of the mCFL, dSC, and 2SC phases.  The low energy excitations 
of these phases are massive or massless gauge bosons, gapped or gapless 
quark quasiparticles, and modes that are associated with fluctuations in
the diquark fields.  Here we concentrate on the gauge bosons and gapless 
fermions.  As for gauge bosons, we only consider the transverse components 
of the gauge fields which, in the static limit, can become massive only 
in a superconductor.  The longitudinal components are Debye screened 
in the vicinity of the transition temperature.  Note that the 
"gapless fermions" in our previous paper \cite{IMTH04} have been defined only 
by diagonalizing the self-energy $\Sigma$ in Eq.\ (\ref{GSig}). In the 
present paper, we have diagonalized the full propagator $G^{-1}(k)$ to 
define the gapless fermions in a more precise manner as will be shown below.

     In Table I, we summarize the symmetries, the number of massive gauge 
bosons, and the number of gapless fermion modes in each phase.  As we will 
show in detail below, the number of gapless quark modes depends not only on 
which phase we consider but also on the temperature itself.  In each phase, 
in addition to the unpaired quark modes whose excitation energies are 
naturally zero, gapless modes appear if the temperature is in the region 
where the gaps satisfy the conditions listed in Table I.
 
     In each phase, the gauge symmetry is broken partially or even totally.
The Meissner masses for the general condensate, Eq.\ (\ref{phase-def}),
and the corresponding massive (or massless) gauge fields can be calculated 
using the GL theory coupled to gluon fields \cite{IB3} as
\begin{eqnarray} 
&&m_{A_1}^2=m_{A_2}^2= \kappa_T g^2 (\Delta_1^2 +\Delta_2^2), \nonumber\\
&&m_{A_4}^2=m_{A_5}^2=\kappa_T g^2 (\Delta_1^2 +\Delta_3^2), \nonumber\\
&&m_{A_6}^2=m_{A_7}^2=\kappa_T g^2 (\Delta_2^2 +\Delta_3^2), 
\label{gluon-masss}\\
&&m_{A_8}^2= \frac{4}{3}\kappa_T g^2 
\left(
\frac{ \Delta_1^2 \Delta_2^2 +\Delta_2^2 \Delta_3^2 +\Delta_3^2 \Delta_1^2 }
{\Delta_1^2+ \Delta_2^2 }
\right), \nonumber\\
&&m_{\tilde{A}}^2=\frac{4}{3}\kappa_T g^2 
\left(
\frac{ \Delta_1^4 +\Delta_2^4 +\Delta_1^2 \Delta_2^2 }
{\Delta_1^2+ \Delta_2^2 }
\right) . \nonumber
\end{eqnarray}
Here $\tilde{A}$ is a mixed field of $A_3$ and $A_8$,
\begin{eqnarray}
\tilde{A}&=&  
\frac{\sqrt 3}{2} 
\frac{\Delta_1^2+ \Delta_2^2 }{\sqrt{ \Delta_1^4 +\Delta_2^4 +\Delta_1^2 
\Delta_2^2 }}A_3 \nonumber\\
&&~~~~~~~~~+\frac{1}{2} \frac{\Delta_1^2-\Delta_2^2 }
{\sqrt{ \Delta_1^4 +\Delta_2^4 +\Delta_1^2 \Delta_2^2 }}A_8.
\end{eqnarray}
Here we followed the notations in Refs.\ \cite{IB, IB3}.  $\kappa_T$ is the 
stiffness parameter which controls the spatial variation of the gap.  In
weak coupling it is proportional to the quartic coupling $\beta$ as \cite{IB3} 
\begin{eqnarray}
\kappa_T=\beta/3 = \frac{7\zeta(3)}{24(\pi T_c)^{2}}N(\mu).
\end{eqnarray}
We can reproduce the results in Table I in Ref.\ \cite{MIHB} by taking a limit
$\Delta_{1,2} \rightarrow  0$ for the isoscalar 2SC phase and 
$\Delta_1=\Delta_2=\Delta_3$ for the CFL phase.  

     As given in a full description in Ref.\ \cite{gapless-AKR}, we can count 
the gapless quark modes by examining the spectra of which the full propagator 
diverges:
\begin{eqnarray} \label{fullD}
{\rm det} G^{-1}(E, {\bf q})= {\cal G} _{Rd,Gu} {\cal G} _{Rs,Bu} 
{\cal G} _{Gs,Bd} {\cal G} _{Ru,Gd,Bs}= 0. \nonumber \\ 
\end{eqnarray}
The gap matrix $\Sigma (k)$ can be written in block-diagonal form if we 
adequately take the base whereas the noninteracting quark propagator $G_0 (k)$ 
is diagonal.  The determinant Eq.\ (\ref{fullD}) can be decomposed 
into three 4 $\times$ 4 determinants ${\cal G}_{Rd,Gu}$, ${\cal G}_{Rs,Bu}$, 
and ${\cal G}_{Gs,Bd}$, and one 6 $\times$ 6 determinant 
${\cal G}_{Ru,Gd,Bs}$, which may effectively be written as
\begin{eqnarray}
\label{det1}
{\cal G} _{Rd,Gu}
&=& \det
 \left(
  \begin{array}{ll} 
  \qs -\hat{p}_{F}^{d}  & -\Delta_3\\
 -\tilde{\Delta}_3&  ~\qs +\hat{p}_{F}^{u}
\end{array}
 \right)  \nonumber\\
  &\times &\det
 \left(
  \begin{array}{ll} 
  \qs -\hat{p}_{F}^{u}  & - \Delta_3 \\
    -\tilde{\Delta}_3& ~\qs +\hat{p}_{F}^{d}  
\end{array}
 \right) ,
  \end{eqnarray}
\begin{eqnarray}
\label{det2}
{\cal G} _{Rs,Bu}
&=& \det
 \left(
  \begin{array}{ll} 
  \qs -\hat{p}_{F}^{s}  & -\Delta_2\\
  -\tilde{\Delta}_2&  ~\qs +\hat{p}_{F}^{u}
\end{array}
 \right)  \nonumber\\
  & \times &\det
 \left(
  \begin{array}{ll} 
  \qs -\hat{p}_{F}^{u}  & -{\Delta}_2\\
     -\tilde{\Delta}_2& ~\qs +\hat{p}_{F}^{s}  
\end{array}
 \right),
  \end{eqnarray}
\begin{eqnarray}
\label{det3}
{\cal G} _{Gs,Bd}
&=& \det
 \left(
  \begin{array}{ll} 
  \qs -\hat{p}_{F}^{s}  & -\Delta_1\\
   -\tilde{\Delta}_1 &  ~\qs +\hat{p}_{F}^{d}
\end{array}
 \right)  \nonumber\\
  & \times &\det
 \left(
  \begin{array}{ll} 
  \qs -\hat{p}_{F}^{d}  & -\Delta_1\\
    -\tilde{\Delta}_1 & ~\qs +\hat{p}_{F}^{s}  
\end{array}
 \right),
  \end{eqnarray}
\begin{eqnarray}
{\cal G} _{Ru, Gd, Bs}
=
 ~~~~~~~~~~~~~~~~~~~~~~~~~~~~~~~~~~~~~~~~~~~~~
~~~~~~~~~~~~~  &&\nonumber\\
 \det
 \left(
 \begin{array}{llllll} 
\qs -\hat{p}_{F}^{u}  & 0& 0  & 0&  \Delta_3  & \Delta_2 \\
0  & \qs -\hat{p}_{F}^{d}&0  & \Delta_3  & 0 &\Delta_1   \\
0& 0&  \qs -\hat{p}_{F}^{s} & \Delta_2  & \Delta_1&0 \\
0&  \tilde{\Delta}_3  & \tilde{\Delta}_2 & \qs +\hat{p}_{F}^{u} &0 & 0 \\
\tilde{\Delta}_3  & 0 &\tilde{\Delta}_1 &0  & \qs +\hat{p}_{F}^{d}&0    \\
\tilde{\Delta}_2  & \tilde{\Delta}_1&0 &0& 0&  \qs +\hat{p}_{F}^{s}  
  \end{array}
 \right). \nonumber \\
\label{det4}
  \end{eqnarray}
Here $\hat{p}_{F}^{i}=p_F^{i} \gamma_0$, and $p_F^{i}$ are the Fermi momenta 
of $i$ quarks.  Only in Eqs.\ (\ref{det1})--(\ref{det4}) we abbreviate 
the Lorentz structure of the gap for simplicity; here $\Delta_i$ indicates 
$\gamma_5 \Lambda^{+} \Delta_i$, and $\tilde{\Delta}_i$ indicates 
$\gamma_0 (\gamma_5 \Lambda^{+} \Delta_i)\gamma_0$.  Moreover, we include the 
effect of $m_s$ as a shift of the effective chemical potential $p_F^i$
following Ref.\ \cite{gapless-AKR}.  Using Eqs.\ (\ref{fermi-mom}), 
(\ref{delmu}), and (\ref{muemus}), $p_F^{i}$'s can be expressed as
\begin{eqnarray}
p_{F}^{u} & =&\mu -(2/3)\mu_e, \nonumber \\
p_{F}^{d}&=&\mu +(1/3)\mu_e,  \\
p_{F}^{s} &=&\mu -(5/3)\mu_e.  \nonumber
\end{eqnarray}
In the vicinity of the critical temperature, chemical potential differences 
between colors $\delta \mu_{ab}$ can be ignored as stated in the previous 
sections.

    As for the three 4 $\times$ 4 determinants, we can count the gapless 
modes as follows.  First of all, ($Rs$,$Bu$) and ($Bu$,$Bd$,$Rs$,$Gs$)
quarks are gapless in the dSC and 2SC phases, respectively, because they do 
not participate in the pairing.  Even for quarks participating in the 
pairing, they can produce gapless excitations near the critical point. 
As an example, consider a Fermi system composed of two species having  
different Fermi momenta, $p_{F}^{(1)}$ and $p_{F}^{(2)}$, and a 
superconducting gap $\Delta$ of pairing with each other.  Then the 
quasiparticle spectra obtained after the diagonalization of the matrix 
fermion propagator can be gapless when the gap is smaller than half of the 
difference of the Fermi momenta,
\begin{eqnarray}  \label{criterion} 
 |p_{F}^{(1)}-p_{F}^{(2)}| > 2 \Delta.
\end{eqnarray}
In Ref.\ \cite{gapless-MS} the detailed analysis of the gapless 
superconductivity at finite temperature has been done; the chemical potentials 
and gaps have been calculated self-consistently at finite temperature. 

     In our case, the difference of the chemical potentials depends on 
temperature only in higher orders so that the left hand side of Eq.\ 
(\ref{criterion}) can be regarded as constant in temperature. 
Then, we can solve Eq.\ (\ref{criterion})
analytically using Eqs.\ (\ref{gap1}), (\ref{gap2}), and (\ref{gap3}).
When the conditions shown below for the mCFL, dSC and 2SC phases
are satisfied, we obtain two more gapless modes in each phase. 
The conditions are
\begin{itemize}
\item{for the mCFL phase, 
$\Delta_2 < \frac{1}{2} \mu_e$. }

\item {for the dSC phase,
$\Delta_1 < \mu_e$. }

\item {for the 2SC phase,
$\Delta_3 < \frac{1}{2} \mu_e$}.  
\end{itemize}
For example, let us consider the temperature satisfying $T_c^{\rm III} < T <  
T_c^{\rm II}$ (the dSC region).  We basically have two unpaired quark modes 
($Bu$ and $Rs$).  Additionally, if the temperature satisfies $\Delta_1 < 
\mu_e $, then we obtain two more gapless modes.

     We move on to the 6 $\times$ 6 determinant.  In this sector, there are 
basically no unpaired quark modes in the dSC and mCFL phases and one unpaired 
quark mode ($Bs$) in the 2SC phase.  In order to find additional gapless modes,
we examine the solutions which satisfy the equation 
\begin{eqnarray}
0&=&{\cal G} _{Ru, Gd, Bs}(E=0,{\bf q})^{1/2}   \nonumber \\
   &=& (|{\bf q}|+ p_{F}^{u})^2(|{\bf q}|+ p_{F}^{d})^2
(|{\bf q}|+ p_{F}^{s})^2 \nonumber \\
&&[\{ x (x-\mu_e) (x+\mu_e) + x \Delta_1^2 +(x- \mu_e) \Delta_2^2 \nonumber \\
 &&~~~~~~~~~~+ (x+\mu_e) \Delta_3^2  \}^2 
+4\Delta_1^2 \Delta_2^2 \Delta_3^2].   
\end{eqnarray} 
Here $x=|{\bf q}|-\bar{\mu}$, where $\bar{\mu}=\mu-2\mu_e/3$ is the chemical 
potential averaged over $Ru$, $Gd$, and $Bs$ quarks.  Then we find 
\begin{itemize}
\item {for the mCFL phase, no more gapless modes.}

\item {for the dSC phase, one more gapless mode corresponding to the
mixture of $Ru$, $Gd$, and $Bs$ quarks. }

\item{for the 2SC phase, if the gap satisfies $\Delta_3 < \frac{1}{2} \mu_e$,
we obtain two more gapless modes corresponding to the mixtures of 
$Ru$ and $Gd$ quarks.}

\end{itemize}

\begin{figure}[t]
\begin{center}
\includegraphics[width=8.5cm]{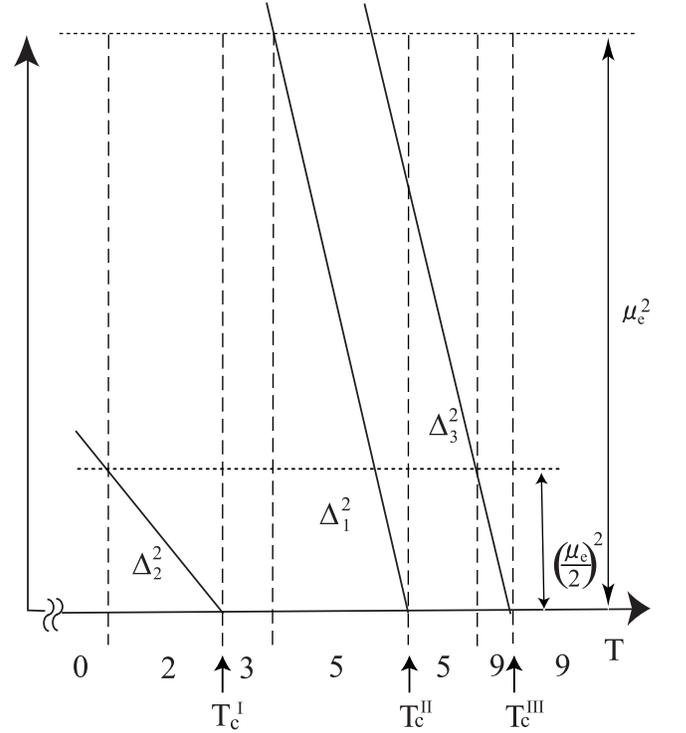}
\end{center}
\vspace{-0.5cm}
\caption{The number of gapless quark modes as a function of temperature,
together with the gaps and the electron chemical potential.}
\label{fig:gapless2}
\end{figure}

     In Fig.\ \ref{fig:gapless2}, we show the number of gapless quark modes
together with the gaps as a function of temperature.  We can see that the 
gapless modes always exist just below the critical temperatures 
$T^{\rm I}_c$, $T^{\rm II}_c$, and $T^{\rm III}_c$.  The number of the 
gapless modes decreases as the temperature lowers at high density.  

\section{Summary and concluding remarks}

     In this paper, we have investigated thermal phase transitions of the 
color superconducting quark matter in the presence of finite strange quark 
mass and under charge neutrality.  We constructed a general form of the GL 
potential in such a way that it fulfills symmetry constraints and then
obtained the weak coupling expression for the GL potential including 
the corrections from unequal quark masses ($m_{i}$, $i$=$u, d, s$) and 
unequal quark chemical potentials ($\mu_{i}$, $i$=$u, d, s$).  The corrections
to the quadratic term of the GL potential from $m_i$ and $\mu_i$ indicate 
that it is the {\em average} density of states of paired quarks which 
determines the critical temperatures of different pairs, as can be seen in 
Eqs.\ ($\ref{eq:final-form-3}$) and ($\ref{eq:final-form-4}$). 

     We found that the effects of the strange quark mass ($m_s$) through the 
mass term in the quark propagator and through the electron charge chemical 
potential play a vital role near the super-normal phase boundary, while the 
color neutrality plays a minor role in contrast to the case at zero 
temperature.
  
     As shown in Figs.\ $\ref{fig:Tc}$ and $\ref{fig:conden}$, an interplay 
between these two effects of $m_s$ splits the single phase transition 
(CFL $\rightarrow$ normal) for the $u$-$d$-$s$ symmetric system to three 
successive second-order phase transitions 
(mCFL $\rightarrow$ dSC $\rightarrow$ 2SC $\rightarrow$ normal).
The window of the "dSC" phase opened up between the 2SC and mCFL phases is 
a new phase where only the superconducting gaps with $d$ quarks are 
non-vanishing \cite{IMTH04}.

     The obtained phase structure, which is shown to be valid in the weak 
coupling (high density) region, may be applicable even in the strong coupling 
region as long as the successive transitions driven by the quadratic term of 
the GL potential take place in a small interval of temperature near the 
super-normal phase boundary.  In connection with this, it was shown recently 
that the dSC phase discussed in the present paper may be replaced by the uSC 
phase in the low density region through a "doubly critical" point on the basis
of the  NJL model \cite{Fukushima} (see also Ref.\ \cite{uSC}).  Moreover, 
if we take into account the chiral condensate at low density,  some interplay 
between the broken chiral symmetry and the color superconductivity is expected
\cite{Bub}.  How to incorporate these aspects in our picture based on the 
GL potential is an interesting open problem.

      In the present paper, the properties of quark and gluon modes
such as the excitation energies were also investigated.  The excitation 
spectra of quarks indicate that in addition to unpaired quark modes, whose 
excitation energy is naturally zero, more than one gapless mode appear just 
below each critical temperature as summarized in Table I.  Effects of the 
gapless modes on physical phenomena such as neutron star cooling is
an interesting problem to be examined \cite{astro}.

      Throughout this paper, we have studied the phase transitions in the 
mean-field level.  In weak coupling, thermal fluctuations of the gluon fields
could change the second-order transition to the first-order one as described 
in Refs.\ \cite{MIHB,GH}.  As far as the order of phase transition is 
concerned, the gluon fluctuations have the following effects (the detailed 
account will be given in our future publication): The second order transition, 
mCFL $\to$ dSC, remains second order.  This is because all eight gluons are 
Meissner screened across $T=T_c^{\rm I}$ and thus cannot induce a cubic term 
in the order parameter in the GL potential.  On the other hand, the 
transitions, dSC $\to$ 2SC and 2SC $\to$ normal, become weak first order 
since some of the gluons become massless at $T=T_c^{\rm I,II}$ (Table I).
In order to obtain a final phase diagram we should study the competition 
between the shift of the critical temperature discussed in the present paper 
and that coming from the fluctuation effects.

\begin{table*}[hbt]
\caption{The symmetry, the number of Meissner screened gluons, and the number 
of gapless fermion modes in the mCFL, dSC, and 2SC phases.  The results for
the Meissner masses are given in Eq.\ (\ref{gluon-masss}).  In each phase, if 
the temperature is in the region where the gaps satisfy the conditions listed 
in this table, more gapless modes appear in addition to unpaired quark modes. 
$(\cdots)$ shows the color and flavor of unpaired quarks.  $\{ \cdots \}$ 
shows which combination of quarks makes the gapless modes. }
\begin{tabular}{|c|c|c|c|c|c|c|c|}
\hline
  & Symmetry & Number of  & \multicolumn{5}{|c|}{Number of gapless quark modes}\\
\cline{4-8} 
 &              & massive gluons & unpaired quarks 
&\multicolumn{4}{|c|}{paired quarks}   \\
\cline{5-8} 
 &              & & & $\Delta_3 < \mu_e/2$
 &$\Delta_1 < \mu_e$ &  $\Delta_2 < \mu_e/2$   & no condition\\

\hline \hline
mCFL &$[U(1)]^2 $ &  8 & 0 & -  & -  & 2 $\{ Rs,Bu\}$   &-\\
\hline
dSC   &  $[U(1)]^4 $ & 8 &  2 $(Rs,Bu)$ & - & 2$\{Gs,Bd\}$& - &1$\{Ru,Gd,Bs\}$\\
 \hline
2SC   & $[SU(2)]^2 \times [U(1)]^2 $ & 5 & 
5&  2 $\{ Rd, Gu\}$   & -&-  &-\\
       &                &     &  ($Bu$, $Bd$, $Bs$, $Rs$, $Gs$) 
&  2  $\{ Ru, Gd\}$   &  &  &\\
\hline
\end{tabular}
 \end{table*}

\begin{acknowledgments}
     We are grateful to H. Abuki, G. Baym, K. Fukushima, and M. Huang
for helpful discussions.  K.I., M.T., and T.H. thank the Institute for 
Nuclear Theory, Univ.\ of Washington where a part of this work has been 
completed.  They also thank K. Rajagopal, I. Shovkovy, T. Sch\"{a}fer, and
C. Kouvaris for discussions.  This work was supported in part by RIKEN Special
Postdoctoral Researchers Grant No.\ A12-52010, and by the Grants-in-Aid of the
Japanese Ministry of Education, Culture, Sports, Science, and Technology
(No.~15540254).
\end{acknowledgments}


\end{document}